# Engineering Graphene Nanoribbons via Periodically Embedding Oxygen Atoms

Yan Zhao[1,2,#], Li-Xia Kang[2,#], Yi-Jun Wang[1], Yi Wu[1], Guang-Yan Xing[2], Shi-Wen Li[2], Jinliang Pan[3], Nie-Wei Wang[4], Yin-Ti Ren[4], Ying Wang[2], Ya-Cheng Zhu[2], Xing-Qiang Shi[4,*], Mengxi Liu[3,5,*], Xiaohui Qiu[3,5,*], Pei-Nian Liu[1,2,*], and Deng-Yuan Li[1,*]

[1]State Key Laboratory of Natural Medicines, School of Pharmacy, China Pharmaceutical University, 211198 Nanjing, China

[2]School of Chemistry and Molecular Engineering, East China University of Science & Technology, 200237 Shanghai, China

[3]CAS Key Laboratory of Standardization and Measurement for Nanotechnology National Center for Nanoscience and Technology, Beijing 100190, China

[4]College of Physics Science and Technology, Hebei University, Baoding 071002, China

[5]University of Chinese Academy of Sciences, Beijing 100049, China

[#]These authors contributed equally.

*Corresponding Authors: shixq20hbu@hbu.edu.cn, liumx@nanoctr.cn, xhqiu@nanoctr.cn, liupn@cpu.edu.cn, dengyuanli@cpu.edu.cn.

# Table of Contents

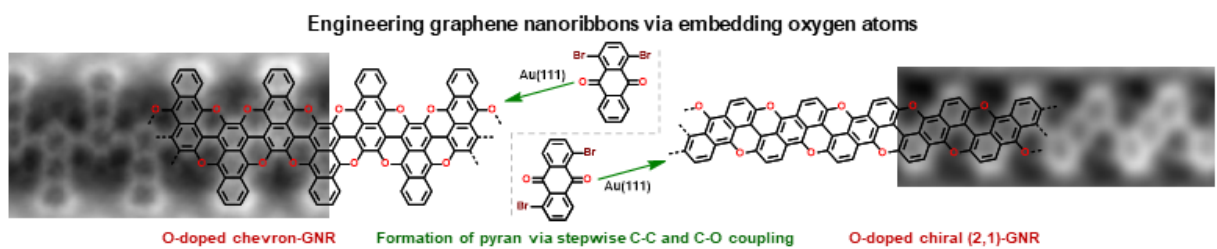

**Abstract**

Heteroatom doping is an important method for engineering graphene nanoribbons (GNRs) because of its ability to modify electronic properties by introducing extra electrons or vacancies. However, precisely integrating oxygen atoms into the lattice of GNRs is unexplored, and the resulting electronic properties remain elusive. Here, we achieve the precise embedding of oxygen atoms into the lattice of GNRs via in situ formation of pyrans, synthesizing two types of oxygen-doped GNRs (O-doped chevron-GNR and O-doped chiral (2,1)-GNR). Using scanning tunneling microscopy, non-contact atomic force microscopy, and density functional theory calculations, the atomic structures and electronic properties of O-doped GNRs are determined, demonstrating that both GNRs are direct bandgap semiconductors with different sensitivities to oxygen dopants. Oxygen dopants have a minor impact on the bandgap of chevron-GNR but a significant effect on the bandgap of chiral (2,1)-GNR, which is attributed to the difference in density of states near the Fermi level between substituted intrinsic carbon atoms and their pristine counterparts. Compared with the pristine chiral (2,1)-GNR, the band structure of O-doped chiral (2,1)-GNR exhibits unexpected band edges transition, which is ascribed to $sp^2$-hybridized oxygen atoms which introduces additional electrons to the conduction band of chiral (2,1)-GNR, leading to the upward shift of Fermi surface.

## Introduction

Graphene nanoribbons (GNRs), which are laterally quantum-confined strips of the honeycomb lattice composed of $sp^2$-hybridized carbon atoms, have risen to prominence as a pivotal material for the potential applications in various technological fields owing to their array of numerous novel physical and electrical properties.[1–8] Their properties are significantly influenced by the boundary conditions, including width,[9–11] edge topology,[12–15] and symmetry.[16,17] These factors can be deliberately tailored through the precise engineering of GNRs.[18,19] The strategic use of substitutional doping with heteroatoms provides extensive opportunities to rationally modulate the physical and catalytic properties as well as the chemical selectivity of GNRs.[20–23] Indeed, the accurate control over dopants within the GNRs' crystal lattice is paramount. The periodic embedding of heteroatoms into the honeycomb lattice of GNRs facilitates the conjugation of the dopants' empty or filled $p_z$ orbitals with the extended π-system, leading to changes in the local electronic environment around the dopant atoms. This process effectively alters the arrangement and bandgap of the energy band structure, thereby introducing dopant states and modifying the local reactivity.[24–26] The advent of "bottom-up" synthesis methods, based on meticulously pre-designed molecular precursors, has marked a significant leap forward in achieving atomic-level precision in the positioning, type, and concentration of dopants within GNRs.[27–30]

So far, typical substitutional atoms of boron,[31–35] nitrogen,[36–42] oxygen,[43,44] and sulfur[44–46] have been achieved to be periodically embedded into GNRs in the form of single doping or co-doping[47–51] along the edge or skeleton. Boron and nitrogen atoms, in particular, have attracted research interest for their roles as electron deficiency and electron donors. As depicted in Figure 1a, graphitic-B with three valence electrons hybridizes $sp^2$ with three adjacent carbon atoms, leaving their $p$ orbital devoid of electrons. This leads to significant hole p-type doping,[52] reducing the nanoribbon's bandgap or generating magnetism upon introducing local magnetic moments.[32] Nitrogen atoms, with five valence electrons, can adopt graphitic-N or pyridinic-N configurations within GNRs. In the case of graphitic-N, the excess electron relative to carbon populates the $p$ orbital, exhibiting electronic characteristics that are complementary to those of boron-doped graphene nanoribbons[21] and potentially generating spin-polarized states.[53] Pyridinic-*N*, positioned at the edge of the nanoribbons, forms bonds with two neighboring carbon atoms through $sp^2$ orbitals, with a single electron in its $p$ orbital and the remaining electrons occupying lower energy hybrid orbitals, only influenced by nitrogen's localized charge, thus retaining electronic properties akin to the undoped system.[53]

Oxygen atoms, distinguished by their smaller atomic radius, higher electronegativity, and electron-rich profile—featuring two additional electrons compared to carbon, could be unique dopants in GNRs. These attributes may lead to lattice distortions, orbital overlap, and local induction effects within GNRs, customizing their electronic properties, enhancing reactivity, and reinforcing structural integrity stability.[54] Dopants may be incorporated into the edge of GNRs through $sp^2$ or $sp^3$ hybridization, resulting in the formation of pyran-O GNRs. In $sp^2$ hybridization (denoted as pyran-O-1), the oxygen atom contributes one electron pair to the GNR π-system and engages other pairs to fill the dangling $sp^2$ orbital, which in turn affects the concentration of free electrons in the nanoribbons. In contrast, oxygen's lone pairs in $sp^3$ hybridization (denoted as pyran-O-2) fill the $sp^3$ orbitals without π-system involvement, potentially leading to electron localization caused by the polarity of oxygen atoms. Although the O-doped chevron-GNRs and OBO-doped chrial (4,1) GNRs on Au(111) have been synthesized using the special precursors with fluorenone[43], furan[44], and OBO group[51] (Figure 1b), the doping patterns adopted by oxygen in GNRs and how they modulate the electrical properties of GNRs remain elusive because of the lack of methodologies for the periodic integration of oxygen into hexagonal rings of GNRs. This limitation hinders in-depth understanding on the hybridization modes of C-O bonds and impedes further development of O-doped carbon materials.

In this study, we have accomplished the periodic embedding of oxygen atoms into hexagonal ring patterns along chevron and chiral GNRs edges. This was achieved by the ingenious introduction of bromine and ketone substituents into the anthracene core (Figure 1c, d), facilitating in situ formation of pyrans through the stepwise debrominative C-C coupling and dehydrogenative C-O coupling reactions. The resulting O-doped chevron GNRs and chiral (2,1)-GNRs have undergone thorough characterization using scanning tunneling microscopy/spectroscopy (STM/STS), non-contact atomic force microscopy (nc-AFM), and density functional theory (DFT) calculations. We found that both types of O-doped GNRs are direct bandgap semiconductors, yet they exhibit varying sensitivities to the effects of oxygen doping. DFT calculations reveal that the contribution of the substituted carbon atoms to the electronic states in the vicinity of the Fermi level can significantly anticipate the degree of impact following oxygen doping. These findings note that the regulatory effect of oxygen doping is intricately linked to the type of nanoribbon and the doping position.

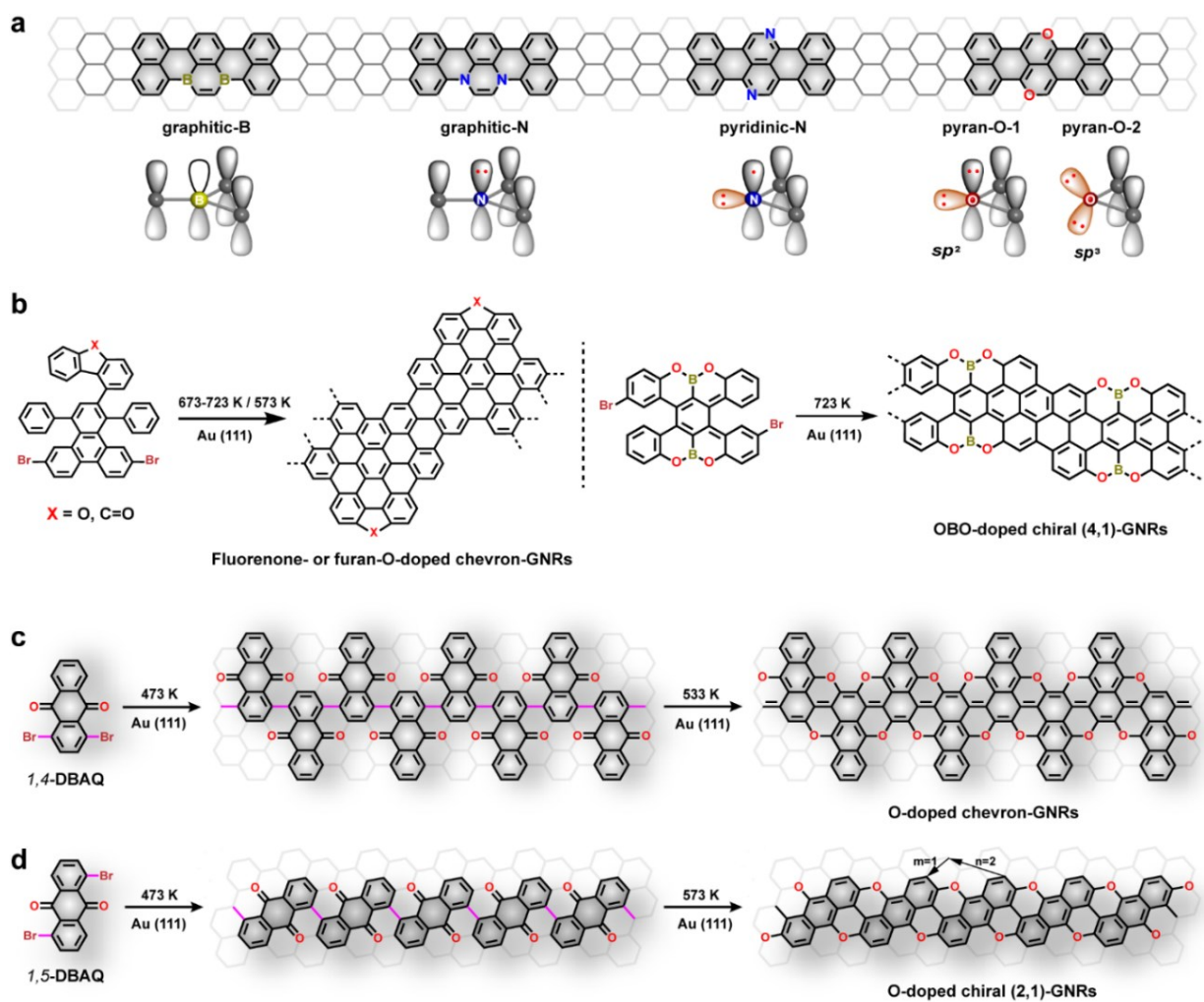

**Figure 1.** Engineering graphene nanoribbons via heteroatom doping. a) Position diagram and electron orbital configuration of graphitic-B, graphitic-N, pyridinic-N, and pyran-O in GNRs. b) Previous reported fluorenone- or furan-O-doped chevron-GNRs (left) and OBO-doped chiral (4,1)-GNR (right). c and d) The synthetic route to prepare pyran-O-doped GNRs on Au (111). The edge geometry of O-doped chiral (2,1)-GNRs is defined by the translation vector (n, m).

## Results and Discussion

**Formation and characterization of O-doped chevron-GNRs**

By harnessing bottom-up synthesis, we have adeptly crafted precisely engineered periodic O-doped GNRs on the Au(111) surface. As shown in Figure 1b, the synthesis of O-doped chevron-GNRs was illustrated using 1,4-dibromo anthraquinone (1,4-DBAQ) as molecular precursors. Two bromine atoms and two ketone groups in 1,4-DBAQ were crucial, serving as the key molecular building block that facilitated the integration of oxygen atoms into the hexagonal rings of GNRs. Our strategy begins with debrominative C-C coupling to form nonplanar 1D polymer chains after the controlled dissociation of bromine groups, followed by the cyclodehydrogenation involving C-O bond formation that embeds oxygen atoms periodically into the GNRs through in situ formation of pyran moieties.

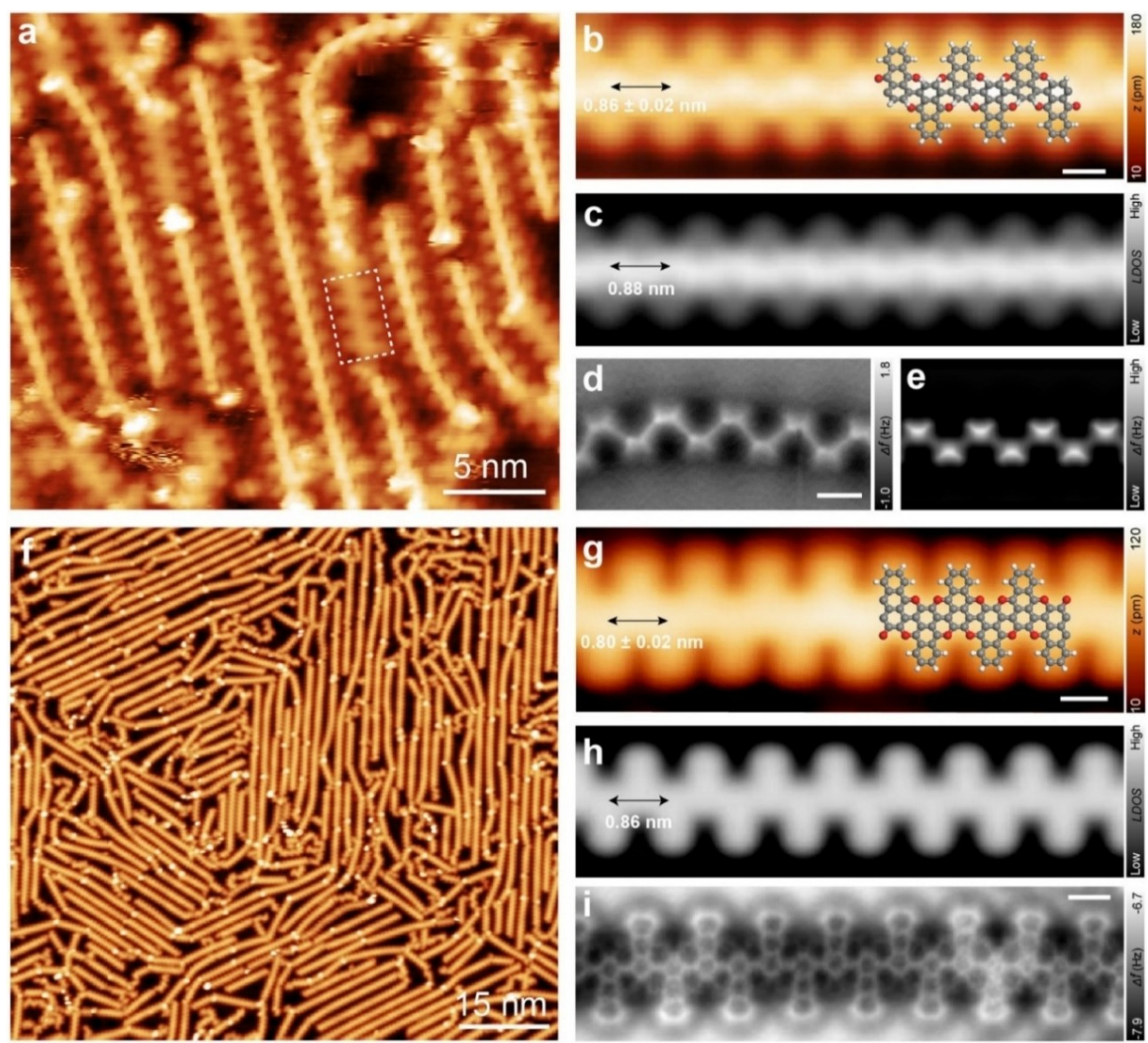

**Figure 2.** Formation of O-doped chevron-GNRs on Au (111). a) Large-scale STM topographic image ($U = -1.50$ V, $I = 530$ pA) after annealed to 473 K. The dashed white rectangle indicated a planar polymer. b) Zoomed-in STM topographic image ($U = -1$ V, $I = 5$ pA, scale bar: 0.5 nm) and c) simulated STM image of the nonplanar polymer. d) Corresponding nc-AFM image (resonant frequency, 32 kHz; oscillation amplitude, 200 pm; scale bar: 0.5 nm) and e) simulated nc-AFM image of the nonplanar polymer. f) Large-scale STM topographic image of O-doped chevron-GNR formed after further annealing at 533 K ($U = -1$ V, $I = 10$ pA). g) High-resolution STM image ($U = -0.35$ V, $I = 80$ pA, scale bar: 0.5 nm) and h) simulated STM image of O-doped chevron-GNR. i) Corresponding nc-AFM image (resonant frequency, 32 kHz; oscillation amplitude, 200 pm; scale bar: 0.5 nm). DFT optimized structure model is placed in b and g, where the red, white, and grey balls represent O, H, and C atoms, respectively.

The 1,4-DBAQ precursor is initially thermally deposited onto the Au(111) surface held at room temperature under ultra-high vacuum (UHV), leading to the self-assembly of 1,4-DBAQ molecules into petal-like structures (Figure S1), stabilized by intermolecular O···H, Br···H and Br···Br interactions. Through a stepwise annealing process, we meticulously monitored the progression of the reaction. At 473 K, the C-Br bonds undergo homolytic cleavage, resulting in 1D polymers through debrominative C-C coupling (Figure 2a). The non-planar structure of the polymeric chains, with protrusions alternating on both sides of the chain axis at a periodicity of approximately $0.86 \pm 0.02$ nm (Figure 2b), aligns with the simulated STM image of 0.88 nm (Figure 2c). The steric hindrance between hydrogen and oxygen atoms of adjacent anthraquinone subunits leads to a rotation about the C-C bonds, causing a divergence in the opposite tilt of successive units on the metal surface. The bond-resolved nc-AFM imaging (Figure 2d) confirms the chain's structure, with the alternating tilt of the subunits matching the simulated nc-AFM protrusions (Figure 2e) along with the polymer, signifying successful polymerization. At certain segments, bright protrusions disappear, reducing the apparent height by approximately 0.06 nm (Figure S2).

Upon annealing to 533 K, the nonplarnar polymer chains underwent the planarization to form the desired O-doped chevron-GNRs with different lengths (Figures 2f and S3), where the periodicity is about $0.80 \pm 0.02$ nm (Figure 2g), as evidenced by the STM simulation (Figure 2h). Nc-AFM measurements have elucidated the occurrence of the pyran annulation, allowing for the distinct localization of carbon and oxygen atoms (Figures 2i and S4). Oxygen atoms are visibly darker against the carbon framework, and their periodic distribution along the chain axis is mapped.[44,55] The edge phenyl rings show slight variations from those in the center, confirming the atomic precision of the hydrogen-terminated armchair edges of the O-doped chevron-GNR (Figure 2i).[56] Additionally, we observed bright features at the junction of two O-doped chevron-GNRs that stand out more than the nanoribbon's structure itself. These features can be attributed to the disruption of the interdimer dehydrogenative C-O coupling step caused by precursor molecules aligned in the same direction during the C-C coupling reaction (Figure S5).

Next, we turn our attention to the local electronic structure of O-doped chevron-GNRs on Au(111), as characterized by STS and DFT calculations. Figure 3a presents the differential conductance (d$I$/d$V$) spectra of O-doped chevron-GNR, taken at the locations marked by colored circles in the STM topography image (inset, Figure 3a), with the spectra of the Au(111) surface state used as a reference. The d$I$/d$V$ spectra over a broad range exhibit pronounced peaks at approximately -0.6 V and 1.5 V relative to the Fermi energy (V = 0 eV). The low-energy dI/dV spectra reveal two additional weak peaks at approximately -0.12 V and 0.10 V, which may be attributed to the valence band (VB) edge and the conduction band (CB) edge of the O-doped chevron-GNR, respectively, with an associated electronic bandgap of 0.22 eV (Figure 3a, insect, and Figure S6a). DFT calculations employing the Heyd-Scuseria-Ernzerhof (HSE06) and Perdew-Burke-Ernzerhof (PBE) functional estimate the bandgap of O-doped chevron-GNR to be 0.76 eV (Figure 3b) and 0.36 eV (Figure S7a), respectively. This appears inconsistent with the experimental results, where the bandgap is significantly reduced upon adsorption on Au(111).

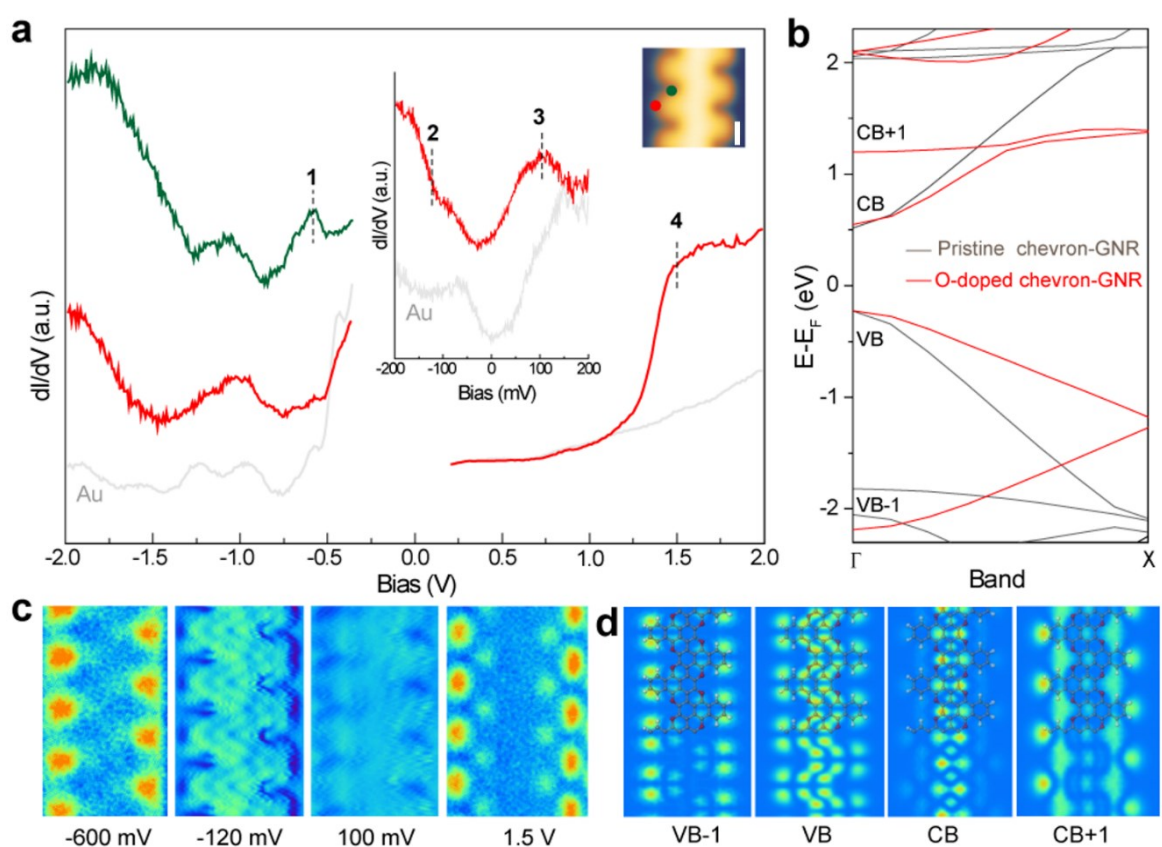

**Figure 3.** Electronic properties of O-doped chevron-GNR on Au(111). a) The d$I$/d$V$ spectra acquired at locations highlighted with red and green markers in the inset STM image (U = -1.0 V, I = 10 pA, scale bar: 0.5 nm). The spectra are shifted for clarity. The grey curves correspond to the reference spectra obtained on a bare Au(111) surface and serve as a marker for the quality of the tip. b) DFT-calculated band structure of the O-doped chevron-GNR and pristine chevron-GNR. c) Constant-current d$I$/d$V$ maps recorded at the indicated bias. d) Calculated local electron density distribution of VB−1, VB, CB, and CB+1 at the Γ point of O-doped chevron-GNRs.

We further performed PBE calculations to analyze the adsorption geometry and charge transfer behavior of O-doped chevron-GNR on the Au(111) surface (Figure S7b). The results revealed that the O-doped chevron-GNR physically adsorbs 3.4 Å above the Au(111) surface due to weak van der Waals interactions, rather than strong chemical bonding. Each unit cell of the O-doped chevron-GNR loses 0.47 electrons to the Au(111) substrate, which could be attributed to the interface push-back effect caused by Pauli repulsion between the GNR's $\pi$-electrons and the Au(111) surface states.[57,58] As a result, the O-doped chevron-GNR exhibits electron-losing behavior, which leads to the upward shift of the original VB and a reduction in the bandgap (Figure S7b).

To avoid the topographic effect on the chain edge of dI/dV maps, we performed the dI/dV mapping in the constant-current mode (Figure 3c and S6b), facilitating a rational comparison with the calculated local electron density distribution (Figure 3d). We found that the states at 1.5 V and -0.6 V are mainly located on the benzene rings at the edge of the O-doped chevron-GNR. In contrast, the state at approximately -0.12 V ( in the range from -120 meV to -50 meV) is clearly localized along the skeleton of O-doped chevron-GNR. These features resemble the local electron density distribution maps of the CB+1, VB-1 and VB (Figure 3d). Unfortunately, we cannot precisely determine the location of CB state of O-doped chevron-GNR because the signals recorded in the range of 85 meV to 1 V were too weak to distinguish the features on Au(111).

DFT calculations reveal slight changes in bond lengths around the oxygen atoms of free-standing chevron-GNRs with and without O dopants (Tables S1 and S2) and the direct bandgap of the O-doped chevron-GNR (0.76 eV) barely differs from that of the pristine carbon-based chevron-GNR (0.72 eV) by a mere 0.04 eV (Figure 3b). When referenced to the vacuum level, the CB edge and VB edge of the O-doped chevron-GNR exhibit similar energy alignments to their pristine counterparts. This is in contrast to GNR doped with oxygen in the planar five-membered rings, where the bandgap is smaller and the CB edge is significantly lower than the pristine GNR.[44] Additionally, we calculated the corresponding projected band structure to explore the impact of oxygen doping on the electronic contribution of the nanoribbon (Figure S8). In the O-doped chevron-GNR, the electronic structure below the Fermi level is primarily provided by the $p_z$ orbitals of the oxygen atoms, while the contribution from oxygen atoms above the Fermi level is relatively weak, with the skeletal carbon atoms playing the dominant role. The possible reason is that each oxygen atom undergoes $sp^2$ hybridization, placing one lone pair in the out-of-plane $p_z$ orbital, thereby adding an extra electron to the entire $\pi$-system. Since the replaced carbon atoms at the edges of the nanoribbon itself contribute little to the nanoribbon states (Figure S8a), the substitution of oxygen atoms does not cause significant changes. This indicates that the doping position of oxygen atoms in the nanoribbon has a pivotal effect on the electronic structure of the nanoribbon.

**Mechanism for the formation of O-doped chevron-GNRs**

For a deeper understanding of the synthetic process, Figure 4 illustrates the proposed three-step reaction pathway for the construction of O-doped chevron-GNRs, rationalized DFT calculations. Initially, the strategically designed bromine atoms are catalytically dissociated by the Au(111) surface, generating surface-bound aryl radical intermediates. The calculated energy barrier from IS1 (initial state) to Int1 (intermediate state) for this initial step is reasonable 0.74 eV (Figure 4a).[59] Upon thermal activation by annealing at 473 K, the debrominated intermediate, Int1, gains sufficient thermal energy to diffuse across the Au(111) surface, forming covalent C-C bonds between monomers through radical coupling. This process results in the formation of a linear polymer chain with a distinct chemical functional pattern, as indicated by the reaction barrier of 0.91 eV (from IS2 to Int2 in Figure 4b), aligning with the initiation of the most energetically demanding C-C coupling.

Delving into the subsequent evolution, we focus on the dehydrogenative C–O coupling, a pivotal step for the annulation to form pyran rings into the nanoribbons. We modelled an infinitely long chain with two molecular units to reveal the atomic details of the reaction. We explored possible reaction pathways from three aspects: the mode of dehydrogenative C–O coupling, the orientation of the H atoms at the reactive sites after coupling (toward or away from the metal surface), and the mechanism of hydrogen elimination (adsorption on the substrate or formation of $H_2$ molecules to leave the system) (Figures 4c and S9-S12). By comparing the reaction energy barriers of different pathways, we identified the optimal path as pathway 1 in Figure 4c: dehydrogenative C–O coupling first occurs between adjacent anthraquinone subunits on the same side of the polymer axis, with positive cooperativity,[60] subsequently propagating unit-by-unit in a domino-like manner[61] until one side is fully coupling, and the other side begins to repeat the process, ultimately completing the entire reaction iteratively.

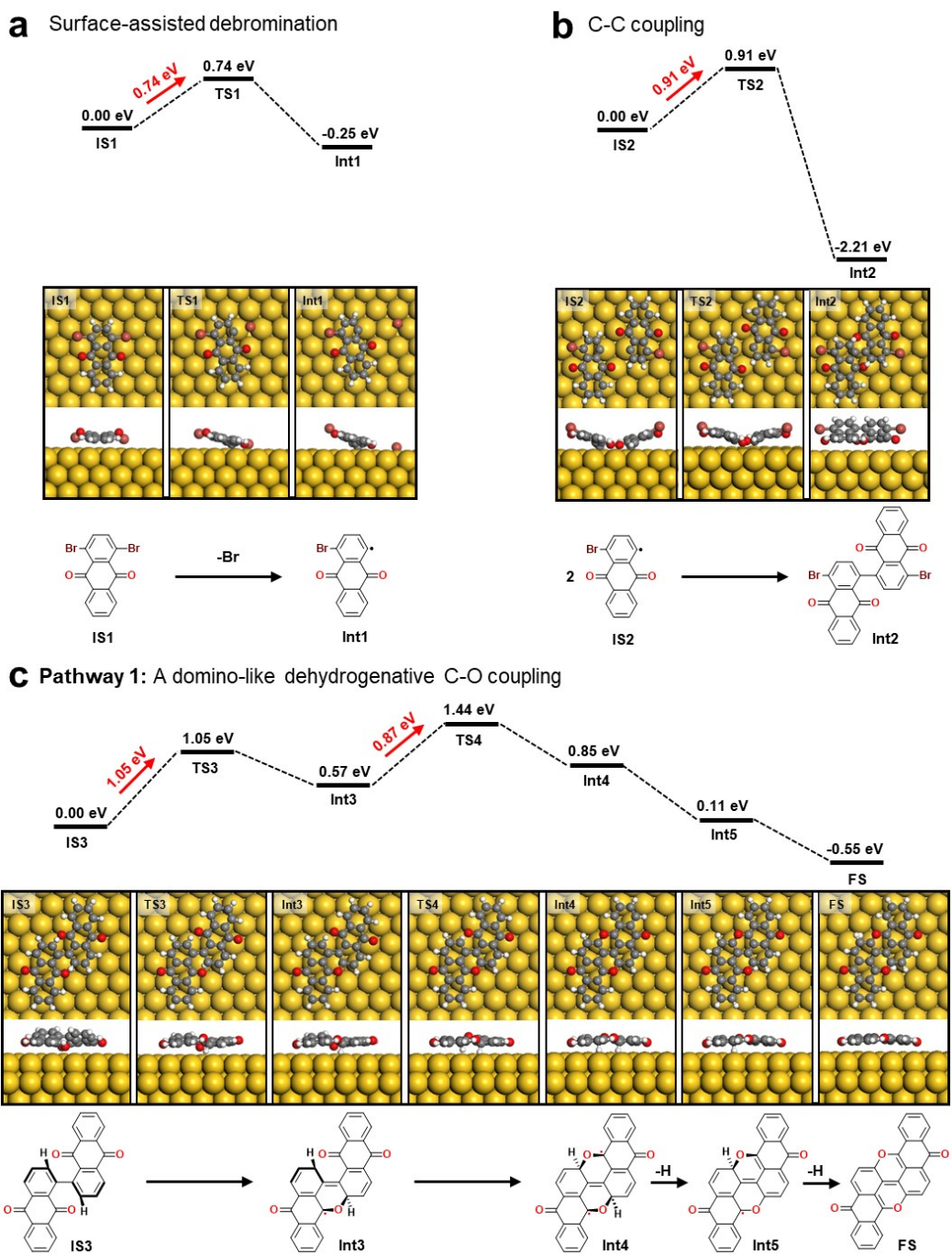

**Figure 4.** DFT-calculated reaction pathway for forming O-doped chevron-GNR on Au(111). a) Surface-assisted debromination to generate aryl radicals stabilized by Au(111). b) C-C coupling of two Au(111) surface-stabilized aryl radicals. c) A domino-like dehydrogenative C–O coupling. The corresponding structure models including top and side views of the initial states (IS), transition states (TS), intermediate states (Int), final states (FS), and chemical structures along the reaction pathway are shown below the energy levels. Small grey, white, and red spheres represent the carbon/hydrogen/oxygen atoms of the structure models, respectively. Large yellow spheres represent Au atoms of the Au(111) substrate.

Initially, driven by the van der Waals interaction between the polymer chain and the metal substrate, one side of the anthraquinone subunits in the polymer chain (IS3) approaches each other, shortening the distance between the same-side aryl carbon and the ketone group, leading to annulation and the formation of a single C–O bond, resulting in Int3. During this process, the H atoms at the reactive sites on the anthraquinone subunits gradually orient toward the substrate. This step has a reaction barrier of 1.05 eV. Subsequently, when all adjacent anthraquinone subunits on the same side of the polymer axis have reacted, the opposing anthraquinone subunits then mirror this process, creating Int4. The initial reaction's preliminary flattening significantly reduces the reaction barrier to 0.87 eV, which is highly favorable for the reaction to proceed smoothly. This observation aligns with the experimental findings that no transition states, similar to those observed in undoped GNRs, were detected.[61] Then, under the catalytic effect of the Au(111) surface, the downward-facing single-sided H atoms are eliminated, forming Int5, and the other side repeats the H elimination process, ultimately yielding the target product FS. In this case, the dehydrogenation proceeds almost spontaneously once the C–O annulation is formed.[62] It is worth noting that the migration of H atoms on aryl carbon is not involved in the whole reaction path. Judging from the entire reaction energy diagram, the highest barrier is the thermodynamically unstable transition state TS4 formed by C–O coupling on one side, which is only 1.44 eV. Moreover, we calculated the rate constant k using the Arrhenius equation[63]: $k = A \cdot e^{-E_a/RT}$, where $E_a$ = 33.09 kcal/mol = 138.45 kJ/mol, $T$ = 533 K, $R$ = 8.314 J/(mol·K), $A$ = $10^{12}$ s$^{-1}$. The resulting k is approximately 0.022 s$^{-1}$, which would imply one such event every 45 seconds. A qualitative comparison with the actual annealing time (10 minutes) active site demonstrated the feasibility of overcoming the 1.44 eV reaction barrier at 533 K. These results indicate that under suitable temperature conditions, the dehydrogenation C–O coupling reaction in a domino-like manner is thermodynamically favorable to form O-doped chevron-GNRs.

**Formation and characterization of O-doped chiral (2,1)-GNRs**

To showcase the versatility of the above-established method and its applicability to a range of GNR geometries for electronic property engineering, we have fabricated O-doped chiral (2,1)-GNRs (Figure 1d). Utilizing 1,5-dibromoanthraquinone (1,5-DBAQ) as a precursor, we have integrated oxygen atoms into the hexagonal ring patterns of chiral GNRs via in situ formation of pyran rings, yielding O-doped chiral (2,1)-GNRs (Figure S13). Figure 5a presents a representative STM topographic image of chiral-shaped GNRs aligned laterally on the Au(111) surface. Zoom-in STM image (Figure 5b) and high-resolution nc-AFM image (Figure 5d) confirm that the gentle stepwise annealing process reliably promotes the formation of C−C and C−O bonds in adjacent anthraquinone subunits, giving rise to O-doped chiral (2,1)-GNRs with pyran rings along their concaved edges. The intramolecular spacing of the fully planarized O-doped chiral (2,1)-GNR is determined to be 0.64 ± 0.02 nm, corroborated by DFT simulation (Figure 5c).

The local electronic structure of O-doped chiral (2,1)-GNR was investigated through site-resolved d$I$/d$V$ spectra acquired at various points above the GNR on Au(111). Figure 5e displays spectra with prominent peaks at -0.05 V and 1.57 V, measured at the oxygen and carbon edges of the nanoribbons, respectively, though with varying peak intensities. The d$I$/d$V$ maps at these energies reveal the spatial distribution of the wave functions associated with the electronic states. The d$I$/d$V$ map at -0.05 V shows the highest intensity localized along the convex carbon edges, while at 1.57 V, the patterns appear at the depressions corresponding to the O-atom locations (Figure 5g). We attribute these two peaks to the valence band (VB) edge and conduction band (CB) edge, respectively, and deduce an electronic bandgap of 1.62 eV for the O-doped chiral (2,1)-GNR.

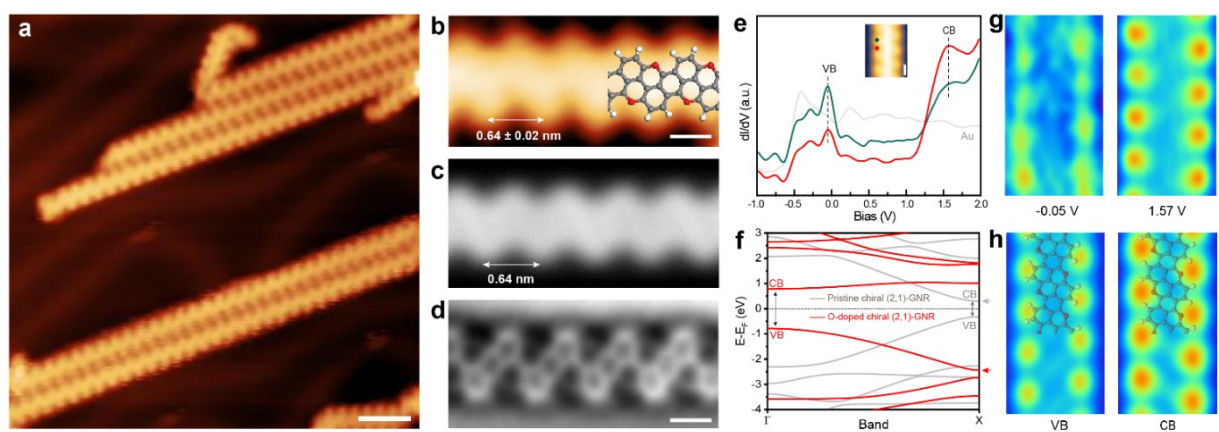

**Figure 5.** Formation and characterization of O-doped chiral (2,1)-GNR on Au(111). a) Large-scale STM image ($U$ = -1.50 V, $I$ = 0.53 nA, scale bar: 5 nm) of O-doped chiral (2,1)-GNR. b) High-resolution ($U$ = 1.58 V, $I$ = 0.49 nA, scale bar: 5 nm) and c) simulated STM images and d) corresponding nc-AFM images (resonant frequency: 40.7 KHz, oscillation amplitude: 100 pm, scale bar: 0.5 nm) of O-doped chiral (2,1)-GNR. e) The d$I$/d$V$ spectra acquired at locations marked by the colored dots in the inset STM image ($U$ = 1.58 mV, $I$ = 0.49 nA, scale bar: 0.5 nm). f) DFT-calculated band structure of O-doped chiral (2,1)-GNR and pristine chiral (2,1)-GNR. g) Constant-current d$I$/d$V$ maps recorded at the indicated biases and h) calculated local electron density distribution of VB and CB at the Γ point of O-doped chiral (2,1)-GNR.

We calculated the band structure for free-standing chiral (2,1)-GNR and O-doped chiral (2,1)-GNR (Figure 5f) to explore the impact of oxygen incorporation on the electronic properties. Notably, the slight changes (similar to chevron-GNR) in bond lengths around the oxygen atoms in chiral (2,1)-GNRs with and without O dopants are also observed in Tables S3 and S4. The calculation estimates a direct bandgap of 1.56 eV for O-doped chiral (2,1)-GNR, which aligns well with the experimental STS findings. The simulated LDOS also maps closely resemble the experimental features (Figure 5h). Compared with the pristine chiral (2,1)-GNR with a direct bandgap of 0.6 eV, the bandgap of O-doped chiral (2,1)-GNR (1.56 eV) is significantly different by approximately 1 eV, exhibiting an unexpected band edges (VBM and CBM) transition from the X point to the Γ point in the Brillouin zone. The calculated projected band structure (Figure S14) indicates that $sp^2$ hybridization of the oxygen atoms introduces additional electrons into the CB of the chiral (2,1)-GNR. This causes an upward shift of the Fermi surface, resulting in the opening of a new bandgap and subversive changes in the positions of band states. In addition, we found that oxygen atoms, similar to replaced carbon atoms in the pristine all-carbon nanoribbon, contribute more to the density of states near the Fermi level. These results demonstrate that the doping effect of oxygen atoms on the electronic properties of GNRs is related to the contribution of the substituted carbon atoms to the original all-carbon GNR.

**Conclusion**

In summary, we have achieved the synthesis of two types of oxygen-doped GNRs (O-doped chevron-GNR and O-doped chiral (2,1)-GNR) on Au(111) via in situ formation of pyrans, where the oxygen atoms were precisely embedded into the lattice of GNRs. Using STM and nc-AFM, we determined the chemical structures and identified the oxygen doping sites in the GNRs. STS measurements combined with DFT calculations demonstrate that both O-doped GNRs are direct bandgap semiconductors with different sensitivities to oxygen dopants. Oxygen dopants have a minor impact on the bandgap of chevron-GNR (from 0.72 eV to 0.76 eV) but a significant effect on the bandgap of chiral (2,1)-GNR (from 0.6 eV to 1.56 eV), which is attributed to the difference in the density of states near the Fermi level between the substituted intrinsic carbon atoms and their pristine counterparts. Notably, compared with the pristine chiral (2,1)-GNR, the band structure of O-doped chiral (2,1)-GNR exhibits an unexpected band edges transition from the X point to the Γ point in the Brillouin zone. It is ascribed to $sp^2$-hybridized oxygen atoms which introduces additional electrons into the CB of the chiral (2,1)-

GNR, leading to the upward shift of the Fermi surface. Our findings provide new insights into understanding the effects of heteroatom doping on the electronic properties of GNRs, which has implications for precisely tuning the electron structure of carbon-based nanomaterials.

## Acknowledgements

This work was supported by the National Natural Science Foundation of China (Nos. 22272050, 21925201, 22161160319, and 22302068), the Shanghai Municipal Science and Technology Qi Ming Xing Project (No. 22QA1403000), the Shanghai Sailing Program (23YF1408700), China Postdoctoral Science Foundation (2023M731080), the China National Postdoctoral Program for Innovative Talents (BX20230126), Shanghai Post-doctoral Excellence program (2023742), and the Fundamental Research Funds for the Central Universities. This work was also supported by the National Key R&D Program of China (No. 2024YFA1208200), Ministry of Science and Technology of China (No. 2017YFA0205000), the National Natural Science Foundation of China (Nos. 22372048, 12274111, 21790353, and 21425310), the Youth Innovation Promotion Association of CAS (No. 2022038) and the CAS Project for Young Scientists in Basic Research (No. YSBR-054).

## Author contributions

D.-Y.L. conceived the project; D.-Y.L., P.-N.L., and X.Q. supervised the experiments; Y.Z., L.-X.K., G.-Y.X., S.-W.L., J.P., and Y.-C.Z. performed on-surface synthesis and scanning probe microscopy experiments; L.-X.K., N.-W.W. and Y.-T. R. conducted theoretical computations with the supervision of D.-Y.L. and X.-Q.S.; D.-Y.L. designed the molecules; Y.-J.W., Y.W., and Y.W. synthesized the molecular precursors. Y.Z., L.-X.K., M.L., and D.-Y.L. analyzed the data. Y.Z. and D.-Y.L. wrote the manuscript. All authors discussed the results and helped write the manuscript at all stages.